\documentclass{IEEEtran}

\usepackage{authblk}
\usepackage{tabularx}

\usepackage{mathtools}
\usepackage{tikz}
\usepackage{multirow}
\usepackage{algpseudocode}  
\usepackage[T1]{fontenc}
\usepackage{array}
\usepackage{booktabs}
\usepackage{latexsym}
\usepackage{caption}
\usepackage{subcaption}

\usepackage{enumitem}
\usepackage{url}

\graphicspath{{./figures/}}
\usepackage{times}

\PassOptionsToPackage{hyphens}{url}\usepackage{hyperref}

\begin{document}

\title{Understanding Content Placement Strategies in Smartrouter-based Peer CDN for Video Streaming}

\author[1]{Ming Ma}
\author[2]{Zhi Wang}
\author[1]{Ke Su}
\author[1]{Lifeng Sun}
\affil[1]{Tsinghua National Laboratory for Information Science and Technology \authorcr Department of Computer Science and Technology, Tsinghua University }
\affil[2]{Graduate School at Shenzhen, Tsinghua University}
\affil[ ]{\{mm13@mails., wangzhi@sz., suk14@mails., sunlf@\}tsinghua.edu.cn}

\maketitle

\begin{abstract}

	Recent years have witnessed a new video delivery paradigm: smartrouter-based peer video content delivery network, which is enabled by smartrouters deployed at users' homes. ChinaCache (one of the largest CDN providers in China) and Youku (a video provider using smartrouters to assist video delivery) announced their cooperation in 2015, to create a new paradigm of content delivery based on householders' network resources \cite{youkuchinacache}. This new paradigm is different from the conventional peer-to-peer (P2P) approach, because millions of dedicated smartrouters are operated by the centralized video service providers in a coordinative manner. Thus it is intriguing to study the content placement strategies used in a smartrouter-based content delivery system, as well as its potential impact on the content delivery ecosystem. In this paper, we carry out measurement studies of Youku's peer video CDN, who has deployed over $300$K smartrouter devices for its video delivery. In our measurement studies, $104$K videos were investigated and $4$TB traffic has been analyzed, over controlled smartrouter nodes and players. Our measurement insights are as follows. First, a global content replication strategy is essential for the peer CDN systems. Second, such peer CDN deployment itself can form an effective sub-system for end-to-end QoS monitoring, which can be used for fine-grained request redirection (e.g., user-level) and content replication. We also show our analysis on the performance limitations and propose potential improvements to the peer CDN systems.
	
\end{abstract}

\section{Introduction} \label{section:introduction}

Video streaming has already become the largest Internet traffic category \cite{Sandvine, cisco}. To meet the skyrocketing growth of bandwidth requirement from the data-intensive video streaming and reduce the monetary cost for renting expensive resource in conventional content delivery networks (CDNs), video service providers are deploying their \emph{peer CDNs} to make use of network and storage resources at individuals' homes for content delivery. Youku, one of the largest online video providers in China, has deployed over $300$K smartrouters at their users' homes in less than one year \cite{youku1million}, expecting to turn a large fraction of its users ($250$M) to such content delivery peer nodes. 

\begin{figure}[!t]
	\centering
		\includegraphics[width=\linewidth]{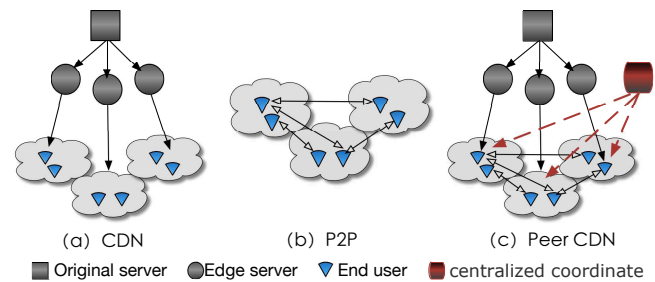}
		\vspace{-0.3cm}
		\caption{The system architecture for CDN, P2P, and peer CDN.}
	\label{fig:CDN_P2P_PCDN}
	\vspace{-0.2cm}
\end{figure}

In Fig.~\ref{fig:CDN_P2P_PCDN}, we plot today's content delivery paradigms including conventional CDN, peer-to-peer (P2P) and peer CDN. Compared to the conventional CDN approach, peer CDN employs network resources which are much closer to users; While compared to the conventional P2P paradigm where users individually cache and serve each other, the nodes in a peer CDN are closely coordinated by the centralized knowledge. For example, the content providers schedule nodes in a peer CDN to proactively cache content, and redirect users to download content from particular smartrouters. Today, traditional content providers and such peer CDN providers even start collaborating to change the traditional content delivery paradigm, to satisfy the ever increasing generated content and edge network requests \cite{youkuchinacache}.

 Such content delivery paradigm can fundamentally change the deployment of content services and applications, and even the roles of individuals play in the Internet, e.g., any individual can become not only a content publisher, but also the corresponding content source hosting the service for the content generated by herself. It is thus intriguing to investigate the details of the peer CDN paradigm, including its performance, strategies used in today's systems, and limitations.


%

In this paper, we conduct extensive measurement to study the peer video CDN of Youku, which has attracted over $300$K users to deploy the smartrouters (namely Youku Router) with $8$GB storage at their homes and offices to serve Youku clients. The challenges to study the system performance and key strategies are as follows: First, the system is distributed in nature, and it is difficult to have a global knowledge using only limited scale of measurement experiments; Second, because the system is a combination of conventional CDN servers and peering nodes, it is challenging to identify the target peer nodes we are interested in; Third, \emph{incentive} mechanisms are used in the system, e.g., users are paid according to the amount of video content they uploaded to other clients, which lead to even more noise in our traffic analysis.

To address these challenges, we design \emph{active} and \emph{passive} measurement experiments to study not only its architecture, but also the key strategies. On one hand, we deploy controlled \emph{clients} in different networks (e.g., CERNET, ChinaUnicom, etc.)~to actively interact with the peer CDN nodes, and study their behaviors; On the other hand, we deploy controlled \emph{smartrouters} in different cities and ISPs to passively observe how these nodes are scheduled by the whole system, and how they serve other clients. Our contributions are summarized as follows.

$\rhd$ To the best of our knowledge, we are the first to conduct measurement to study the real-world smartrouter-based peer video CDN systems. Based on our measurement studies covering $104$K videos, $132$ conventional centralized CDN servers and $4$M users, we present not only the architecture and protocols, but also the key content placement strategies of smartrouters that can significantly affect the performance of the peer CDN system.
			 
$\rhd$ We reveal that global content replication is essential for the peer CDN paradigm for the on-demand video streaming service, and proactively scheduling replication and caching on a daily basis is used by Youku smartrouters, e.g., the average lifespan of chunks is $24.2$ hours. 

$\rhd$ We observe some interesting features of videos that affect the content placement strategies, including the video popularity, the video freshness and the system recommendation. To name a few, the popularity distribution of videos cached by smartrouters follows the zipf-like distribution, $70\%$ of the reality shows cached by our smartrouters are downloaded within $7$ days after their release, and $73\%$ of the videos cached by smartrouters are recommended on the fontpage of the Youku website.

$\rhd$ We also explore the limitations and propose the potential improvements for today's video peer CDN systems.



The rest of the paper is organized as follows. In Sec.~\ref{sec:measure}, we introduce our measurement scheme. The video peer CDN architecture is presented in Sec.~\ref{section:architecture}. The smartrouter-level video placement strategies are analyzed in Sec.~\ref{section:chunk}. We present related works in Sec.~\ref{section:relatedwork} and conclude the paper in Sec.~\ref{section:conclusion}.

\section{Measurement Methodology}  \label{sec:measure}

We conduct both passive and active measurements on smartrouters and clients in our study.

$\rhd$ Passive Measurement by Controlled Smartrouters. We use $5$ Youku smartrouters in our measurement study, which are deployed in different locations with different ISPs. The $5$ routers are denoted as follows: \begin{itemize}
\setlength{\itemsep}{0pt}
\setlength{\parsep}{0pt}
\setlength{\parskip}{0pt}
\item \textsf{R1},\textsf{R2}, which are deployed in Beijing (ISP: CERNET), with public IP addresses and $80$ Mbps downlink/uplink bandwidth; 

\item \textsf{R3}, which is deployed in Beijing (ISP: CERNET), with private IP (NAT) and $80$ Mbps downlink/uplink bandwidth; 

\item \textsf{R4}, which is deployed in Beijing (ISP: China Unicom), with public IP and $2$ Mbps downlink bandwidth and $512$ Kbps uplink bandwidth; 

\item \textsf{R5}, which is deployed in Shenzhen (ISP: CERNET), with private IP address and $80$ Mbps downlink/uplink bandwidth.

\end{itemize}


Our testing nodes are the ordinary version of Youku routers and our monitoring period is from Nov 1st to Dec 30th, 2015. The details of our passive measurements are as follows:

1) File System Monitoring: By performing root injection\footnote{http://openwrt.io/docs/youku/}, we are able to login these devices via SSH, and monitor the file systems on the devices, where \emph{video files} (chunks and video files are used exchangeably in this paper) are stored to serve users. In Youku peer CDN, a content is cached as a video chunk with an unique content ID, and our experiments cover $104$K different chunks. In Sec.~\ref{section:chunk}, we will present the details of the file monitoring results.

2) Traffic Monitoring: We monitor the traffic patterns on these devices, using the conventional network utility tools including \textsf{tcpdump}, \textsf{netstat}, etc. Tab.~\ref{tab:datasets} illustrates the statistics in our measurement experiments. For instance, the dataset contains $132$ unique Youku peer CDN servers and $4$M unique IPs.

\begin{table}[!t]\footnotesize
  \caption{Statistics in our measurement studies.}
  \label{tab:datasets}
  \centering  
    \begin{tabular}{ll}
    \toprule
     Time period		&	$11/1 - 12/30, 2015$	\\
    \midrule 
     Video number	    & $104,380$ \\
     Total traffic 	& $4.7$TB \\
     TCP/UDP Packets				&	$7,251,347,984$	\\
	 Contacted servers	    &	$132$		\\
	 Contacted IPs				&  	$4,207,836$\\
	 Distinct ASes			&  $4,015$\\
    \bottomrule
  \end{tabular}
  \vspace{-0.3cm}
\end{table}
 
$\rhd$ Active Measurement by Controlled Clients. In our study, we also run ordinary Youku clients to join the system, watch a number of videos like ordinary users, and measure how their requests are served by other smartrouters and traditional CDN servers.

%
%


\section{Architecture and System Workflow Inferred} \label{section:architecture}

Based on our measurement results, we infer the architecture used by the Youku smartrouter-based CDN.

\begin{figure}[!t]
	\centering
 		\includegraphics[width=0.85\linewidth]{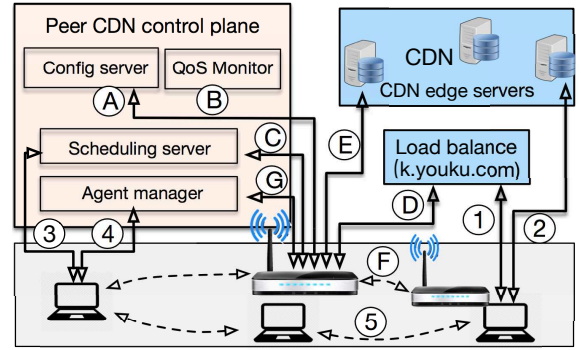}
		\caption{Architecture of the peer CDN system.}
   		\label{fig:architecture}
   		\vspace{-0.6cm}
\end{figure}

Fig.~\ref{fig:architecture} illustrates the general architecture of the Youku peer CDN system. There are three basic components in Youku peer CDN.
 
1) \textbf{Agents}: There are three types of agents in the system, including:

 $\rhd$ \emph{Peer Routers}: Dedicated smartrouters which can cache and distribute contents;
 
  $\rhd$ \emph{Peer clients}: Clients who install the Youku Accelerator and cache videos watched to serve others; 
  
  $\rhd$ \emph{Ordinary users}: Clients who only watch videos on Youku.

2) \textbf{Peer CDN control servers}: This component generates the primary scheduling strategy, e.g., which videos should be replicated by which peer routers. There are mainly $4$ types of servers as follows:

	$\rhd$ Config servers: Peer routers download configuration parameters from Config servers (\textsf{A}). In our traffic monitoring, we have traced the HTTP response in the configuration XML file, e.g., ``\url{param name=`pull-hot-resources-timeval' value=3600}'' indicates that peer routers will download chunks with the cycle of $3600$s, and we will confirm this in the next section.

	$\rhd$ QoS monitors: Peer routers report statistics (\textsf{B}) to the QoS monitor server, including the information of their partners and their operation states, e.g., the download and upload speeds of partners will be reported. A peer CDN can have a global views of end-to-end QoS using the monitoring mechanism.
		
	$\rhd$ Scheduling servers: Scheduling servers schedule the content replication (\textsf{C}) according to the information monitored. First, smartrouters receive replication tasks from the scheduling servers to download new contents to their local storage; Second, smartrouters are redirected to CDN edge servers or other peer routers, to actually download these chunks. 
	
	$\rhd$ Agent managers: A set of agent managers are deployed in different ISPs and locations. Each of them manages the smartrouters that are close to it, e.g., in the same ISP and location. Agent managers also help clients to find smartrouters from which they can download the video content.
		 
		 	 
3) \textbf{CDN Infrastructure}: The CDN infrastructure takes two types of responsibilities. The first one is to serve users as regular CDN servers, when users cannot download from smartrouters; The second responsibility is to publish video contents for peer video CDN, by pushing the latest content to the peer routers (\textsf{D},\textsf{E}). Youku equally segments a video into different chunks (usually with a duration of $6$ minutes), to be cached by the smartrouters. The average chunk size is $17$MB; hence one Youku smartrouter with the $8$GB TF card can cache up to $400$ chunks.

In this peer CDN system, HTTP protocol is used to download contents from edge servers, and a private P2P protocol based on UDP, is adopted for the peers to delivery their data. A smartrouter downloads the content from multiple peers in parallel (\textsf{F}) when it obtains the peer list. Once the peer router caches a content, it is added to a chunk swarm (a subset of routers who cache the same video content) and ready to serve users or other peer routers. When the user starts to watch a video, s/he will follow the steps from step \textsf{1} to step \textsf{5}. The user may request the first video chunks from the conventional CDN servers (\textsf{1},\textsf{2}) to start playing as soon as possible. At the same time, s/he asks the scheduling server whether to switch to the peer CDN service mode, and the steps to activate the peer CDN service are \textsf{3},\textsf{4},\textsf{5}, which are similar to \textsf{C},\textsf{G},\textsf{F}, respectively.

\section{Content Placement Strategies Measured} \label{section:chunk}

We investigate the content deployment strategies adopted in the smartrouter-based peer CDN, which have a great impact on the system overall performance. In this section, we study how content replication is triggered on smartrouters, what factors affect the content replication, and limitations of current strategies.

%

\begin{figure*}[htbp]
	\begin{minipage}[b]{0.65\linewidth} 
	  		\begin{subfigure}[b]{.49\linewidth} 
				   \includegraphics[width=\linewidth, height=3.9cm]{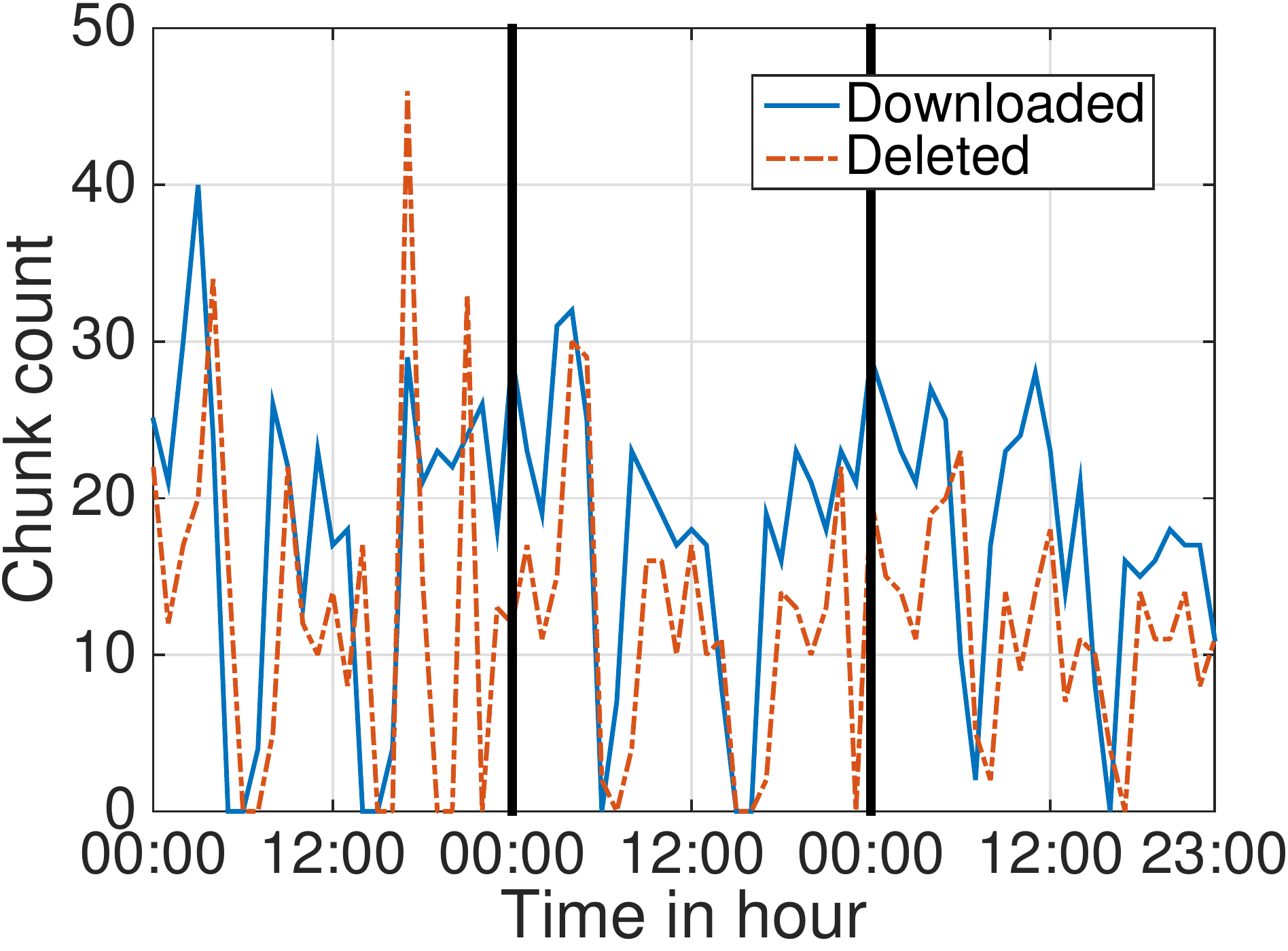}
					\caption{Content replication pattern of \textsf{R1}.}  
	  		\end{subfigure}
	  		\hfill
	  		\begin{subfigure}[b]{.49\linewidth} 
				   \includegraphics[width=\linewidth, height=3.9cm]{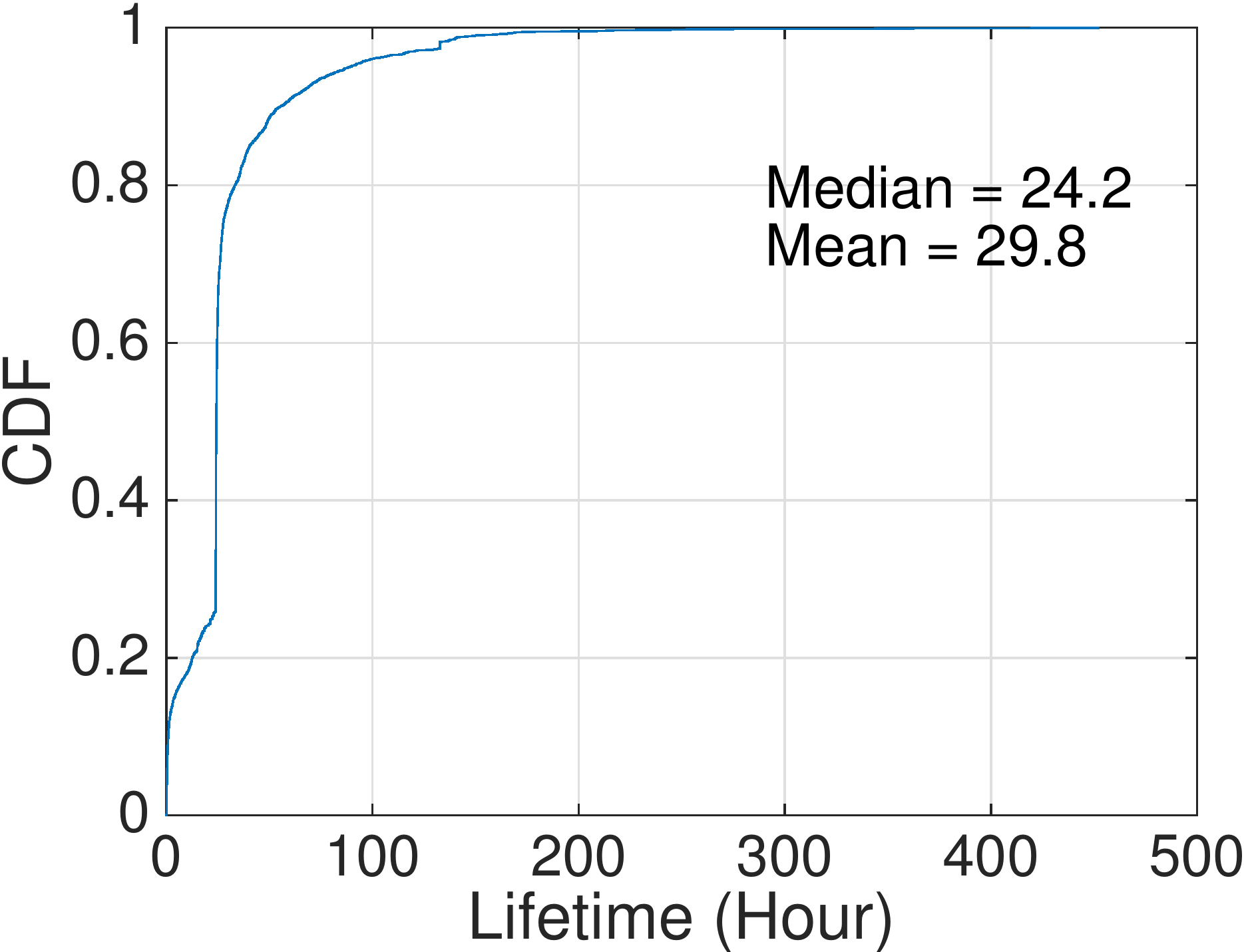}
					\caption{The distribution of the chunk lifespan.}
	  		\end{subfigure}  
      		\caption{The trigger of chunk replication scheduling.}
     		\label{fig:chunkscheduling}  
	\end{minipage}
	\hfill
	\begin{minipage}[b]{0.33\linewidth} 
    \centering
    \includegraphics [width=1\linewidth, height=4cm]{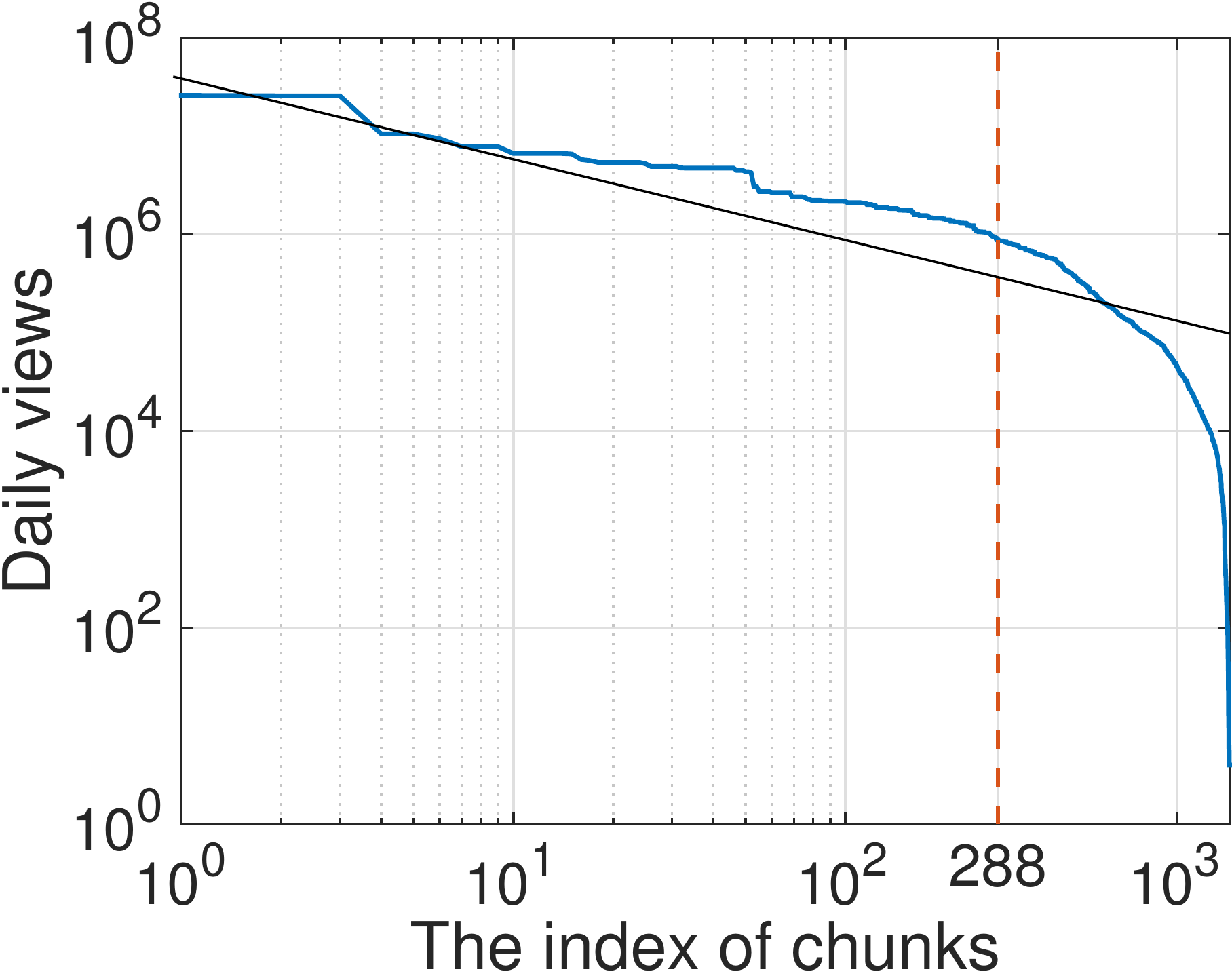}
		\caption{The video popularity distribution.}
		\label{fig:popularity}
  \end{minipage}
  \hfill
  \vspace{-0.2cm}
\end{figure*}

 
%
Note that our 5 smartrouters exhibits similar patterns for download, upload and caching, though the amount of data they download and upload is different due to different networks they are deployed in. Thus we only present the patterns of one router in this paper unless otherwise specified.

\subsection{What Triggers Content Replication}
For smartrouter-based peer CDN, it is essential to explore how centralized controller manages and schedules the content placements among a large amount of peer routers.

\subsubsection{Global Content Replication}

We study the triggers of content replication, by analyzing when chunks are downloaded or deleted over time. In Fig.~\ref{fig:chunkscheduling}(a), we plot the numbers of chunks downloaded to the smartrouter, and removed from the smartrouter in each hour over $3$ days (Other days are similar). The downloaded chunk number follows a daily pattern, i.e., the lower levels of chunk downloaded happen between $5$-$7$AM and $2$-$4$PM every day, and the peaks happen between $0$-$3$AM, which indicates that chunk downloading is scheduled periodically (i.e., hourly) during a day. Moreover, we count that $22\%$ of the chunks are downloaded from Youku edge CDN servers and most of the chunk ($78\%$) placements are accomplished by peer routers themselves, which effectively alleviates the load of Youku servers.

As for the chunk removal, it is also scheduled by the centralized control, instead of only using the cache replacement algorithms (e.g., Least Recently Used algorithm). For example, $46$ chunks are removed during $4$-$5$PM of the first day shown in Fig.~\ref{fig:chunkscheduling}(a) when there is still $30\%$ of the storage capacity available, which indicates Youku maintains \emph{centralized cache information} to determine which content item is cached by which smartrouter.

In order to figure out whether the Youku peer CDN uses a different content placement strategy when the peer routers are using different ISPs or at different locations, we evaluate the chunk similarity between any two peer routers by using Jaccard index \cite{jaccard}, i.e., the size of the intersection of two chunk sets divided by the size of the union of two chunk sets. The results show that, for any two peer routers, the average chunk similarity is between $4.2\%$ to $9.8\%$, which remains at the similar degree, even though the peer routers are deployed in different networks, indicating the smartrouters are scheduled to cache videos using similar strategies.


\subsubsection{Periodical Content Update}

To explore how frequently the contents in the peer routers are updated, we plot the distribution of the chunk lifespan in Fig.~\ref{fig:chunkscheduling}(b). The median (resp. mean) chunk lifespan is $24.2$ hours (resp. $29.8$ hours), i.e., most of the chunks cached in the peer router for about one day. There are plenty of fresh and popular published videos that users want to watch, but the router storage is much smaller than the CDN edge server, which results in a frequent and timely content update in the peer routers. This rate of chunk update is similar to the change rate of top ranked videos measured in the VoD systems \cite{yu2006understanding, PPTVmobile}.




\subsection{What Affects Content Replication}

We examine the characteristics of the chunks downloaded in our smartrouters, to observe what kind of videos are replicated by peer video CDN to meet the users' requests.

\subsubsection{Recent Global Content Popularity}

Video popularity is the most critical influential factor on the video placement. In this part, we explore the correlation between the videos cached in our peer routers and their global popularity.
 
In order to obtain a dataset of video information of chunks downloaded to our peer routers, we sample $1436$ cached chunks on \textsf{R1}, and crawl the popularity of videos of these chunks from the Youku website\footnote{http://index.youku.com/} in one week. We refer to these videos as \emph{peer-videos}, which are in $5$ categories: TV series ($54\%$), variety show ($20\%$), movie ($13\%$), cartoon ($5\%$) and others ($8\%$). We plot the number of daily views of these videos against the rank of these peer-videos in Fig.~\ref{fig:popularity}. We observe that the popularity distribution of the most popular videos cached on our smartrouter follows a zipf-like distribution.




%

Next, we present the relationship between the content replication on smartrouters and the popularity of the videos. In our datasets, we have collected the global most popular $100$, $600$ and $1200$ videos based on their views of the current day, current week, current month and total history on Youku, respectively. In Tab.~\ref{tab:videopop}, we show the proportion of the most popular videos among the peer-videos. Our observations are as follows: 1) Videos that are popular in history are not necessarily cached by smartrouters, e.g., no video cached by our smartrouter is among the $100$ most popular videos in history. The reason is that smartrouters are usually scheduled to only cache videos that are recently popular. 2) The videos that are popular in the recent month are likely to be cached by smartrouters, e.g., over $60\%$ of the peer-videos are among $1200$ most popular videos of the recent month. The reason is that the most popular video genre in China is TV series \cite{PPTVmobile}, which is successively published on a daily basis and a TV series can be popular for around one month.

\begin{table}[htbp]\footnotesize
  \caption{Proportions of the most popular videos replicated by peer routers.}
  \label{tab:videopop}
  \centering  
    \begin{tabular}{llll}
    \toprule
     Top number		&	$100$		&	$600$		&	$1200$			\\
    \midrule 
     Current day 	&	$15.1$\%	&	$18.3$\%	&	$29.0$\%		\\
     Current week	&   $32.1$\%	&	$47.6$\%	&	$52.8$\%		\\
     Current month	& 	$23.4$\%	&	$50.8$\%	&	$61.1$\%		\\
     History			& $0$\%		& 	$8.7$\% 	&	$13.1$\%		\\
    \bottomrule
  \end{tabular}
  \vspace{-0.2cm}
\end{table}
\subsubsection{Local Content Popularity} 

We also study the impact of local content popularity on smartrouter content placement. In particular, we investigate whether the cached contents of a smartrouter are influenced by the videos watched by the local users, who own the smartrouter. At $0$AM on Nov $1$, we sampled $10$ videos from the list of this month's top ranked videos, which had not been cached by our smartrouters yet. By associating our laptops and mobile devices with the Wi-Fi networks provided by \textsf{R1} and \textsf{R4}, we let the Youku clients install on these devices play these videos $50$ times each day for one weeks. We observe that none of them is replicated to the smartrouters during this period. This observation indicates that the Youku content replication strategy is insensitive to the local content popularity, and scheduled globally.

\subsubsection{Video Freshness vs. Video Recommendation}

We study the impact of video freshness and video recommendation on content replication, respectively. We define the video freshness as the number of days between the day when a video is published and the day when this video is replicated to our smartrouters. Video recommendation means the recommendation of videos by Youku, e.g., the frontpage of Youku generally presents the newly published videos, recommended videos and popular video links, which can attract users to click \cite{yu2006understanding}. 

\begin{figure}[htbp]
	
      \begin{minipage}[t]{0.98\linewidth} 
    \centering
        \includegraphics[width=\linewidth]{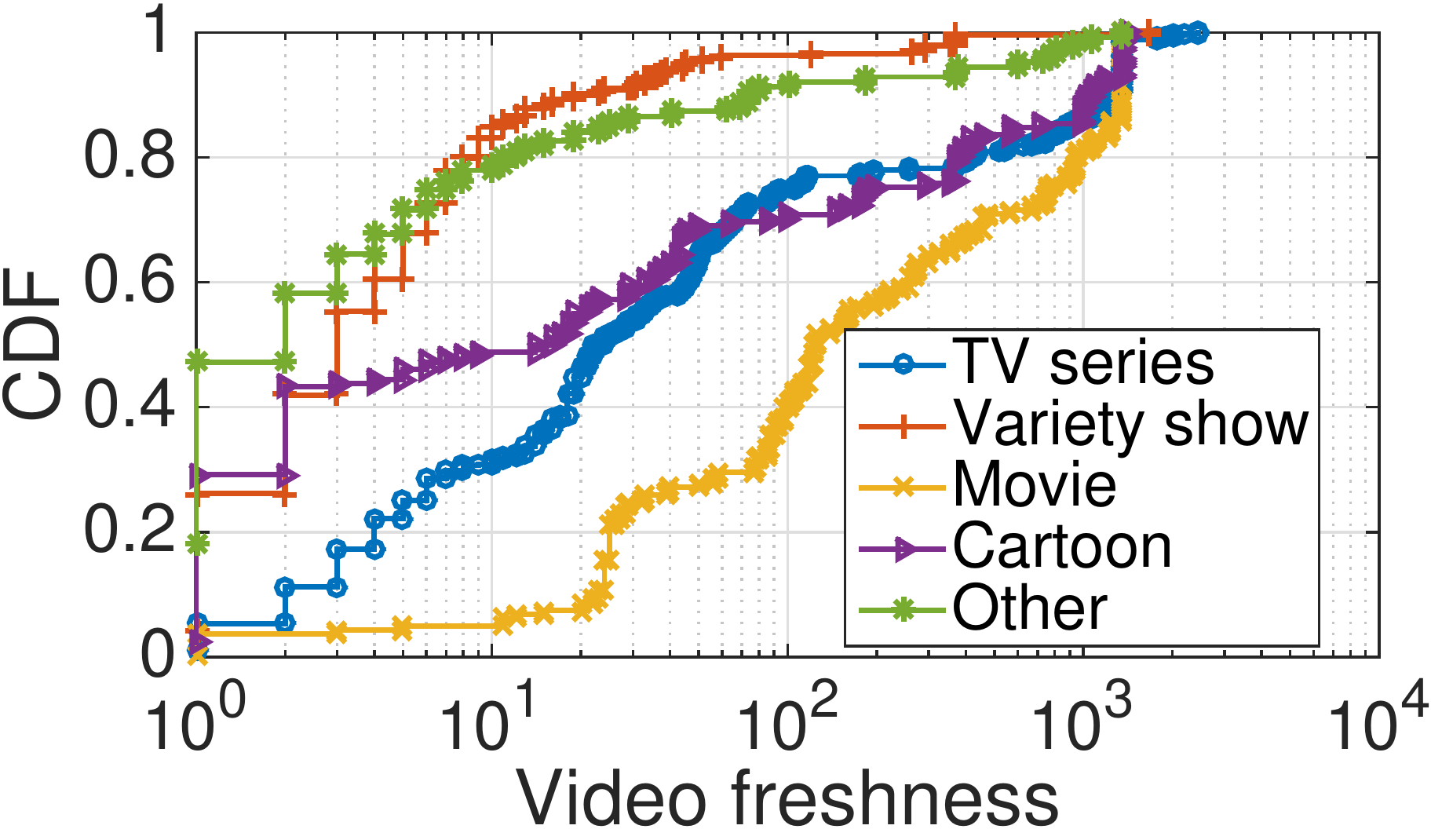}
    \centerline{\scriptsize (a) CDF of freshness levels of videos replicated to routers.}
        \end{minipage}
      \\ \\ \\ 
  \hfill 
  \centering
   \begin{minipage}[t]{0.93\linewidth} 
    \centering
\includegraphics[width=\linewidth]{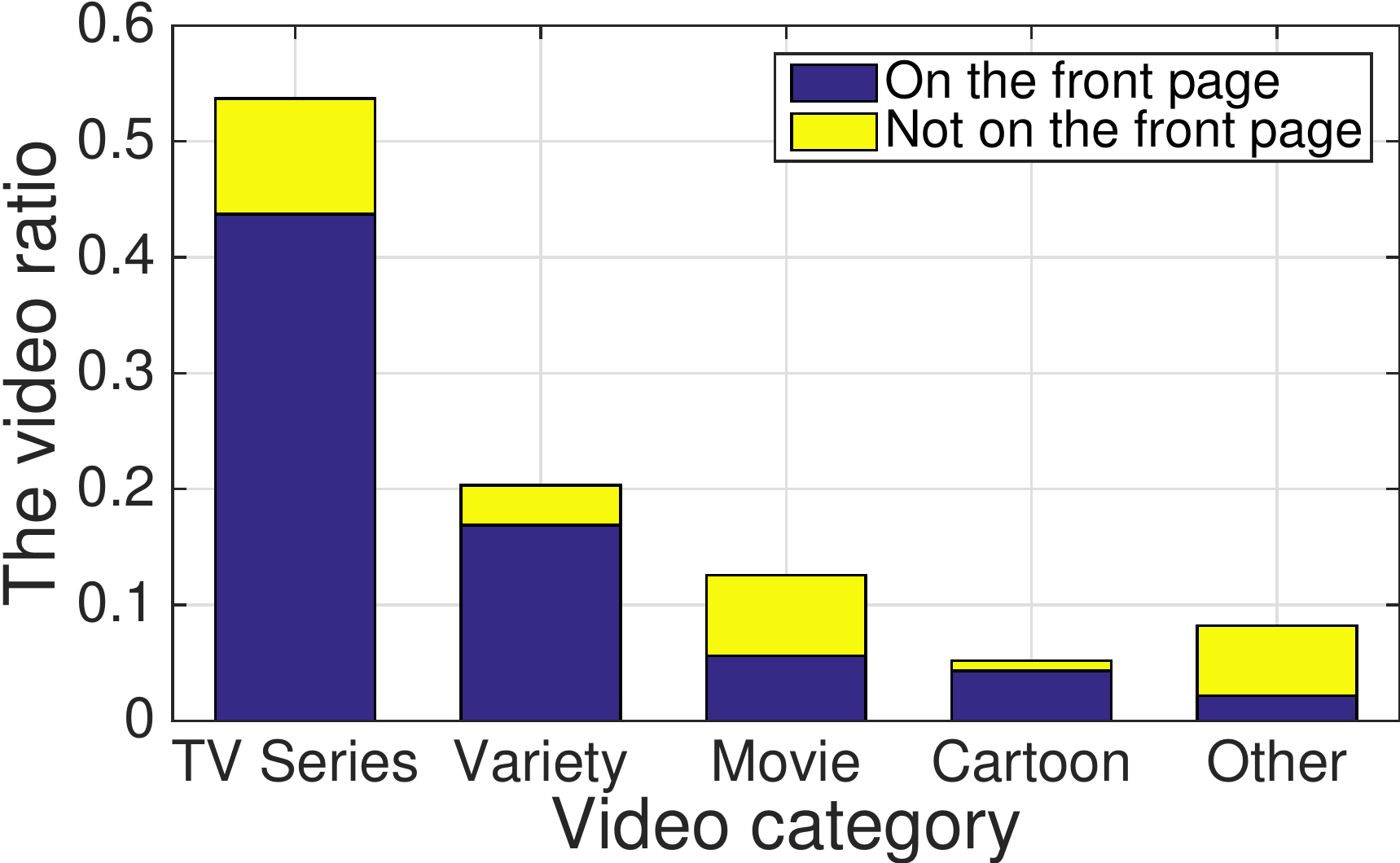}
  \centerline{\scriptsize (b) Fraction of recommended videos replicated to routers.}
     \end{minipage}
  \hfill 
 \caption{Impact of video freshness and recommendation on content placement.}
  \label{fig:chunkcontent}  
  \vspace{-0.25cm}
\end{figure}

We first study the impact of the video freshness on content placement at smartrouters. In Fig.~\ref{fig:chunkcontent}(a), we plot the CDF of video freshness of the videos cached by our smartrouters. We observe that the CDFs differ significantly for different video categories. For example, $50\%$ of movies with freshness larger than $100$ days are still cached by our smartrouters, while only $5\%$ of variety shows with freshness larger than $100$ days are cached. In particular, $70\%$ of the reality shows cached by our smartrouters are downloaded within $7$ days after their release.

Next, we study the impact of video recommendation on content replication. Fig.~\ref{fig:chunkcontent}(b) shows the fraction of peer-videos that are recommended by Youku on the frontpage. We observe that $73\%$ of the videos cached by our peer routers are on the frontpage, and TV series and cartoon have the highest possibility to be cached by smartrouters when they are recommended.
	

\subsection{Limitations of Current Strategies}




We examine the limitations of current content placement strategies used by the peer video CDN of Youku.

\subsubsection{Inefficient Smartrouter Allocation}

We first study whether the number of allocated smartrouters matches the popularity of popular videos. We sample $65$ videos cached by our routers and crawl the list of other smartrouters which are allocated to cache these chunks during $11$ days, by repeatedly requesting peer lists for $1500$ times on \textsf{R1}, \textsf{R4} and \textsf{R5}, using the APIs provided by the agent manager servers.

\begin{figure*}[htbp]\centering
    \begin{minipage}[t]{0.32\linewidth} 
    \centering
  	\includegraphics [width=\linewidth]{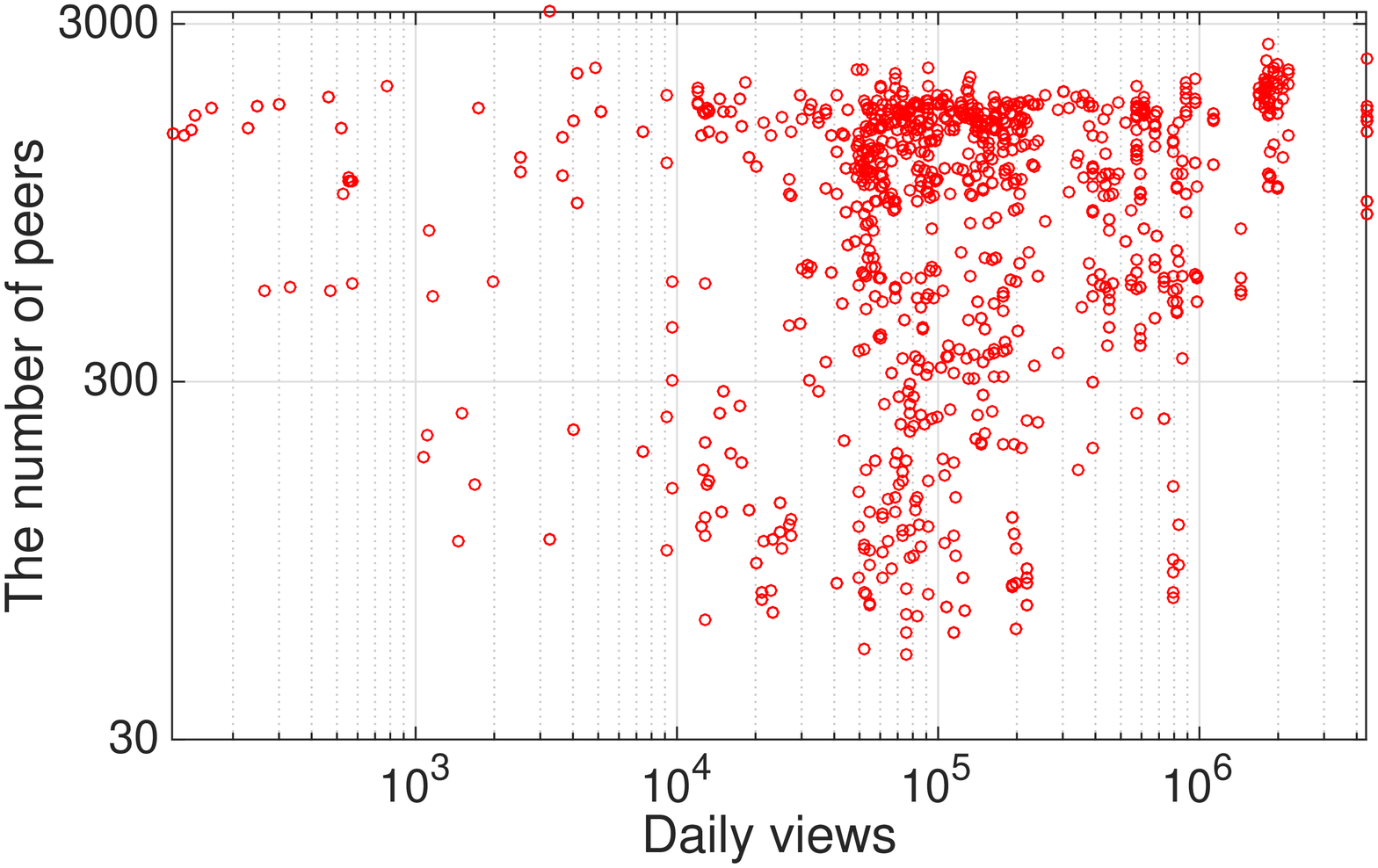}
 		\centerline{\scriptsize (a) Peer number vs. the daily views.}
  \end{minipage}
  \hfill
      \begin{minipage}[t]{0.32\linewidth} 
    \centering
  	\includegraphics [width=\linewidth]{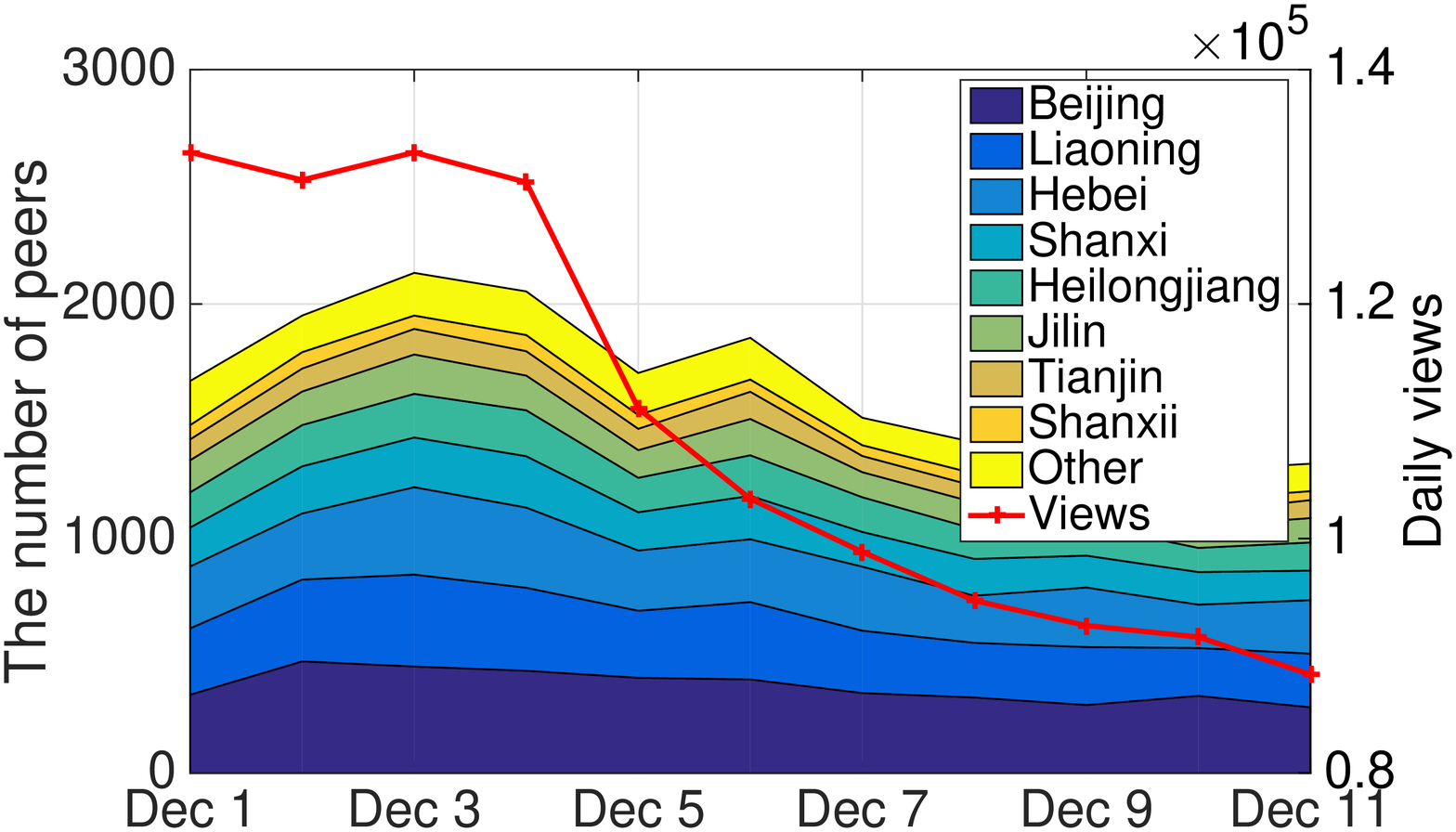}
 		\centerline{\scriptsize (b) Case 1: replication response.}
  \end{minipage}
  \hfill
        \begin{minipage}[t]{0.32\linewidth} 
    \centering
  	\includegraphics [width=\linewidth]{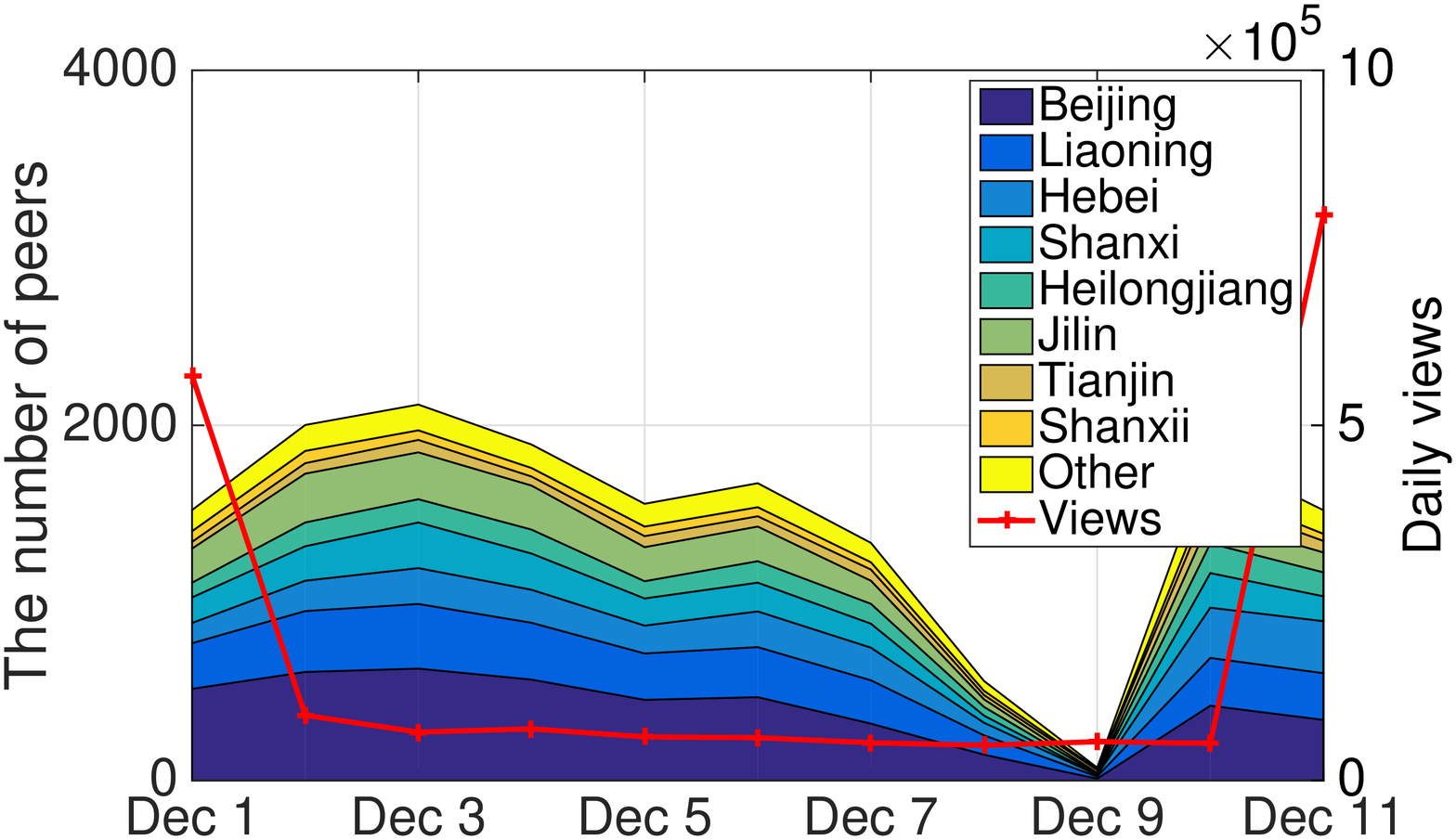}
 		\centerline{\scriptsize (c) Case 2: replication response.}
  \end{minipage}
  \hfill
 \caption{Number of replicas versus video popularity, and replication response.}
  \label{fig:chunk_peernu}
   \vspace{-0.2cm}
 \end{figure*}

Fig.~\ref{fig:chunk_peernu}(a) shows the results crawled from \textsf{R4}: each sample is the number of allocated smartrouters (i.e., number of replicas) for a video versus its popularity (i.e., the number of daily views of the video). It is surprising to notice that there is no obvious correlation. The similar results are also observed on the datasets of other smartrouters in our study. The reasons are as follows: 1) New published videos and recommended videos are deployed more aggressively on smartrouters than popular videos. 2) Current content replication algorithm of Youku fails to efficiently schedule content replication according to the popularity distribution of videos.


\subsubsection{Slow Replication Response}

Next, we study the replication response, by investigating the change of the number of allocated smartrouters for videos. Fig.~\ref{fig:chunk_peernu}(b) and Fig.~\ref{fig:chunk_peernu}(c) illustrate the number of smartrouters hosting two videos over time. The curves of different provinces in China are stacked to show the geographical distribution of the smartrouters. We observe that the changes of the number of allocated smartrouters happen after the changes of the video popularity, e.g., in Fig.~\ref{fig:chunk_peernu}(c), the number of allocated smartrouters drops around one week after the popularity drops from the Dec.~2 to the Dec.~9. This observation indicates that the current content replication algorithm has a relatively large delay for content replication. The video in Fig.~\ref{fig:chunk_peernu}(c) is recommended on the frontpage on Dec.~10, and we observe that many smartrouters seem to be scheduled to cache it before the recommendation.



%


%
%
%

%
\section{Related Work} \label{section:relatedwork}

Video traffic has already represented a significant fraction of today's traffic, e.g., a recent report indicates that Netflix and Youtube account for $55\%$ of the peak downstream traffic in North America by Dec 2015 \cite{Sandvine}, and Cisco predicts that 3/4 of the world's mobile data traffic will be video by 2020 \cite{cisco}. Thus it is important to alleviate the load of both CDN and network backbone. Many researchers realize that the abundant resources, such as set-top \cite{cdn_orchestrate}, small cell base station (SBS) \cite{poularakis2014approximation}, and smartrouter \cite{OfflineDownloading, thunder}, can be utilized to assist the content delivery and offload the traffic of original server.

There are many measurement works studying the content distribution system, including the conventional CDN \cite{ZhanghuiControlPlane}, P2P \cite{PPTVmobile, PPTVmobile2015} and Peer CDN \cite{zhang2015XunleiKanKan, Akamai, YinLiveSky}, which provide valuable insights for network properties, user behaviors, and system dynamics to guide the future development of Peer CDN. In NetSession \cite{Akamai}, peers cache the contents downloaded by end users. LiveSky \cite{YinLiveSky} delivers live streaming video, and its architecture is necessarily different from the peer video CDN in our study. \cite{thunder} studies a smartrouter-based peer CDN system, which mainly provides static file downloading. However, the description of the content placement strategy of this work is relatively simple, i.e., pushing $80$TB traffic per day based on the file popularity in previous day. 

Our previous work \cite{Ming-icccn16} present the detailed system analysis, such as the transport protocols, peer selection strategy and QoS performance. In this paper, we focus on the content placement strategies and its limitations of the real-world smartrouter-based peer video CDN system.

\section{Concluding Remarks} \label{section:conclusion}

Smartrouter-based peer CDN leads to a new content delivery ecosystem, which deploys dedicated router at users' homes to replicate content and serve other users. To understand the performance, limitations and potential improvements of such content delivery systems, in this paper, we conduct a comprehensive measurement on a real smartrouter-based peer CDN platform deployed by Youku, which serves $250$ million users. Our findings show that today's peer CDNs enable prompt global scheduling over millions of peer routers, including pushing newly published videos to smartrouters and dynamically replacing staled content. Such prompt and global strategies are enablers for today's user-generated or fresh video distribution. In particular, our measurement studies reveal that (1) Peer video CDN systems generally adopt global replication and caching strategies; (2) Contents are updated more frequently in smartrouters than in conventional CDN servers; (3) Factors including content popularity and freshness, and recommendation all affect the content placement strategies; (4) There are some limitations of the strategies used by Youku peer CDN, including the slow content replication response.

We also discuss the limitations of our measurement approach: According to the results, our $5$ smartrouters demonstrate similar download/upload/cache patterns. However, it is possible that some smartrouters behave differently under special network conditions. It is still intriguing to study behaviors of all smartrouters from the system operator's perspective.

Finally, we propose the potential improvements to today's smartrouter-based peer video CDNs which are meaningful for future designs: (1) The centralized controller should collect more end-to-end QoS information (e.g., RTT and bandwidth between peers) from the smartrouter swarms, which can be utilized for fine-grained user request redirection; (2) More advanced strategies are still expected, e.g., given the valuable user behaviors, e.g., social relationship \cite{zhi-acmmm2012, zhi-tmm-cpcdn2014}, collected by the system, more sophisticated predication frameworks for video popularity can be designed to improve the content replication responsiveness.

\bibliographystyle{plain}
\bibliography{mylib}

\end{document}